\documentclass[aps,prl,preprint,groupedaddress]{revtex4}

\usepackage{graphicx}

\begin{document}

\title{Kondo excitons in self-assembled quantum dots}

\author{A.~O.~Govorov$^{1,2}$}
\author{K. Karrai$^{3}$}
\author{R. J. Warburton$^{4}$}
\affiliation{$^{1}$Department of Physics and Astronomy, Ohio
University, Athens, Ohio 45701, USA} \affiliation{$^{2}$Institute
of Semiconductor Physics, 630090 Novosibirsk, Russia}
\affiliation{$^{3}$Center for NanoScience and Sektion Physik,
Ludwig-Maximilians-Universit\"{a}t, Geschwister-Scholl-Platz 1,
80539 M\"{u}nchen, Germany} \affiliation{$^{4}$Department of
Physics, Heriot-Watt University, Edinburgh EH14 4AS, UK}

\date{\today}

\begin{abstract}
We describe novel excitons in quantum dots by allowing for an interaction
with a Fermi sea of electrons. We argue that these excitons can be realized very simply with
self-assembled quantum dots, using the wetting layer as host for the Fermi sea.
We show that a tunnel hybridization of a charged exciton with the Fermi sea
leads to two striking effects in the optical spectra. First, the photoluminescence
lines become strongly dependent on the vertical bias. Secondly, if the exciton spin is
nonzero, the Kondo effect leads to peculiar photoluminescence line shapes with a line width
determined by the Kondo temperature.
\end{abstract}

\maketitle

The Kondo effect concerns the interaction of a localized spin with
a Fermi sea of electrons \cite{book}. It is a crucial effect in
various areas of nanophysics where often a nano-sized object with
spin is in close proximity to a metal \cite{Kondo-effect0}.
Semiconductor nanostructures are ideal for investigations of Kondo
physics because, in contrast to metals, their properties are
voltage-tunable \cite{Kondo-effect1}. Characteristic to the Kondo
effect are a screening of the localized spin and a resonance in
the quasi-particle density of states (DOS) at the Fermi energy.
This leads to a maximum in conductance at very low temperatures in
the transport through a spin-$\frac{1}{2}$ quantum dot
\cite{vanderWiel}. So far, the Kondo effect in nanostructures has
been studied almost exclusively in relation to transport
properties. In optics, Kondo effects have only been discussed
theoretically with respect to nonlinear and shake-up processes in
a quantum dot \cite{Raikh,Kikoin}.

We present here the theory of novel effects which arise from a
combination of the Kondo effect and the optics of nanostructures.
In our model, a charged exciton in a quantum dot (QD) interacts
with a Fermi sea of free electrons. We call the resulting
quasi-particles Kondo-excitons. We suggest that such novel
excitons can exist in self-assembled quantum dots where charged
excitons can be prepared in voltage-tunable structures
\cite{Nature}. Furthermore, the wetting layer can be filled with
electrons and used as a two-dimensional (2D) electron gas
(fig.~1a), offering a simple means of generating a Fermi sea in
close proximity to a quantum dot. A significant point is that
self-assembled dots are very small, only a few nanometers in
diameter, allowing the Kondo temperature $T_{K}$ to be as high as
$\sim 10$ K. We find that the photoluminescence (PL) from Kondo
excitons is determined by the Kondo DOS at the Fermi level such
that the PL line width is equal to $k_B T_K$ ($k_B$ is the
Boltzmann constant). Furthermore, the PL energy depends strongly
on the Fermi energy, and therefore also on a bias voltage. This
behavior contrasts with conventional single dot PL where the lines
are very sharp depending only weakly through the Stark effect on
the bias \cite{Nature}.

In voltage-tunable structures, self-assembled QDs are embedded
between two contacts \cite{Nature,APL}. This makes it possible to
control the number of electrons in a QD by application of a
voltage $U_g$ (fig.~1a) \cite{APL}. By generating a hole with
optical excitation, it has been demonstrated that there are
regions of gate voltage with constant excitonic charge, with the
excitonic charge changing abruptly at particular values of $U_g$
\cite{Nature}. The charged excitons, labelled $X^{n-}$, contain
$n+1$ electrons and one hole. Excitons with $n$ up to 3 have been
observed in InAs/GaAs quantum dots \cite{Nature,Zrenner}. At
higher voltages, capacitance-voltage spectroscopy shows that
electrons fill the 2D wetting layer \cite{APL}. In this case, the
Fermi energy ($E_F$) in the wetting layer depends linearly on
$U_g$: we find that $E_F=(a_o^*/4d) (U_g-U_g^o)$, where $a_o^*$ is
the effective Bohr radius, $d$ is the distance between the back
gate and wetting layer, $U_g^o$ is the threshold gate voltage at
which electrons start to fill the wetting layer, and $d\gg a_o^*$
\cite{APL}. Existing PL results in the regime of wetting layer
filling show broadenings and shifts of the PL lines \cite{Nature},
which we believe is some evidence of an interaction with the Fermi
sea, although this has not been analyzed in detail.

\begin{figure}
\includegraphics[angle=0,width=0.7\textwidth]{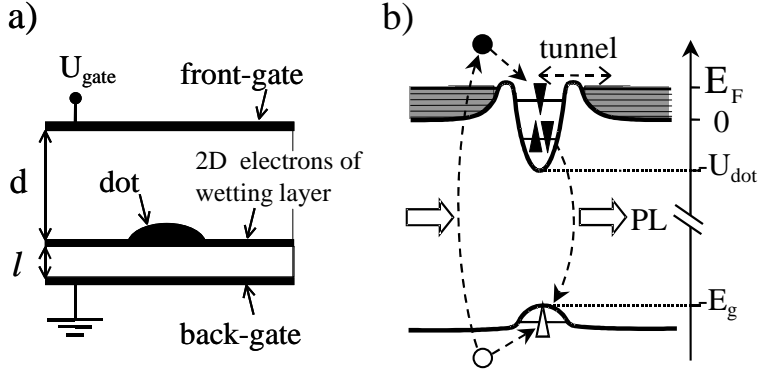}
\caption{\label{fig1} (a) Schematic of the heterostructure with a
quantum dot embedded between front and back gates; $d\ll l$. (b)
The band diagram of a quantum dot and associated wetting layer
showing also the energies typical to self-assembled quantum dots.}
\end{figure}

Typically, the QD $X^{0}$ and $X^{1-}$ excitons are strongly bound and exist only
at gate voltages where the wetting layer is empty. It
is therefore unlikely that $X^{0}$ and $X^{1-}$ can couple with the Fermi
sea. Instead, we focus on the excitons $X^{2-}$ and $X^{3-}$ for which
the electrons on the upper state may couple with extended states of the Fermi sea
as we illustrate in fig.~1b.
Since self-assembled QDs are usually anisotropic, we model a
QD with two bound non-degenerate orbital
states, having indexes $s$ and $p$.
For a $p$ state electron, the electrostatic potential consists of the
QD short-range confinement and the long-range Coulomb repulsion from the charges
in the lower $s$ state.
This potential will therefore have a barrier at the edge of the QD (fig.~1a),
allowing the $p$ electron(s) to tunnel in and out of the QD.

To investigate the interaction between a quantum dot exciton and a
Fermi sea of electrons, we employ the Anderson Hamiltonian which includes the single-particle energy
$\hat{H}_{sp}$, the intra-dot Coulomb interaction, and a
hybridization term $\hat{H}_{tun}$:
\begin{eqnarray}
\hat{H} & = &
\hat{H}_{sp}+\frac{1}{2}\sum_{\alpha_1,\alpha_2,\alpha_3,\alpha_4}
U^{ee}_{\alpha_1,\alpha_2,\alpha_3,\alpha_4}a_{\alpha_1}^+a^+_{\alpha_2}
a_{\alpha_3}a_{\alpha_4}-\sum_{\alpha_1,\alpha_2,\alpha_3,\alpha_4}
U^{eh}_{\alpha_1,\alpha_2,\alpha_3,\alpha_4}a_{\alpha_1}^+b^+_{\alpha_2}
b_{\alpha_3}a_{\alpha_4}+\hat{H}_{tun}, \label{H}
\end{eqnarray}
where $a^+_{\alpha}$ ($b^+_{\alpha}$) is the intra-dot
creation operator of electrons (holes); $U^{ee}$
and $U^{eh}$ are electron-electron and electron-hole Coulomb
potential matrix elements, respectively. The index
$\alpha$ stands for $(\beta,\sigma)$, where $\beta$ is the orbital index and $\sigma $ the
spin index; $\beta$ can be $s$ or $p$,
and $\sigma=\pm \frac{1}{2}$ for electrons and $\pm \frac{3}{2}$ for heavy holes. $\hat{H}_{tun}$ is given by:
$\hat{H}_{tun}=\sum_{\sigma}V_k[
c_{k,\sigma}^+a_{p,\sigma}+a^+_{p,\sigma}c_{k,\sigma}]$,
where the operators $c_{k,\sigma}$ describe the delocalized
electrons in the Fermi sea; $k$ and $E_k$ are the 2D momentum and
kinetic energy, respectively. For the tunnel matrix element $V_k$,
we assume $V_k=V$ in the interval $0<E_k<D$ and $V_k=0$ elsewhere
\cite{D}. In the operator $\hat{H}_{tun}$, we include only
coupling between the $p$ state and the Fermi sea. In our approach,
the quantization in a QD is assumed to be strong so that the
Coulomb interactions can be included with perturbation theory
\cite{GovorovJETP,Richard}. The tunnel broadening is the smallest
energy in the problem. Our approach is to solve eq.\ \ref{H} for
the initial and final states, and calculate the optical emission
spectrum at zero temperature:
\begin{eqnarray}
\nonumber
 I(\omega)=Re \int_0^{\infty}dt e^{-i\omega
t}<i|\hat{V}_o^+(t)\hat{V}_o(0)|i>,
\end{eqnarray}
where $|i>$ is the initial state and the operator
$\hat{{V}_{o}}=V_{opt} (b_{s,-\frac{3}{2}}a_{s,\uparrow}+
b_{s,\frac{3}{2}}a_{s,\downarrow}+{\rm c.c.})$ describes the
strong transitions between the hole and electron $s$ states.

\begin{figure}
\includegraphics[angle=0,width=0.7\textwidth]{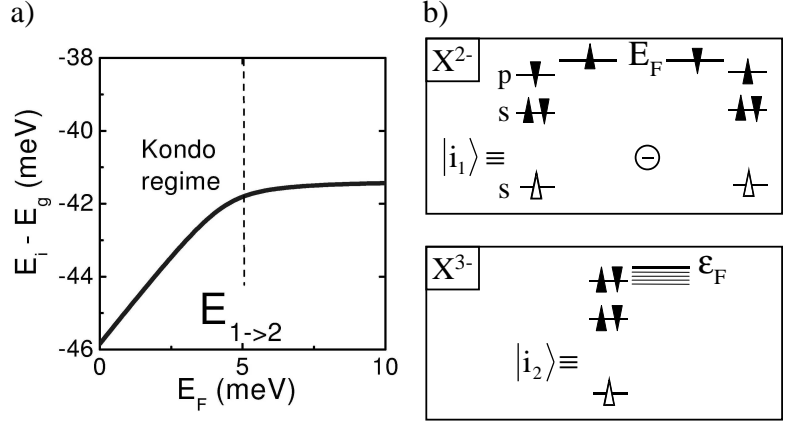}
\caption{\label{fig2} (a) Calculated energy of the initial state
as a function of the delocalized state energy. The energy $E_F$
plays the role of the Fermi level. Here, $U_{dot}=64$ meV and
$E_g$ is the band gap of the QD. (b) Electron configurations
contributing to the initial ground state. Electron (hole) spins
are represented with solid (open) triangles. }
\end{figure}

We start with a zero bandwidth model, in which the Fermi sea is
replaced by a single ``delocalized" state of energy $E_{F}$
\cite{book}. In this simplified model, $E_{F}$ plays the role of
the Fermi level allowing us to predict the main features in the
evolution of the PL with Fermi energy. We describe the electron
wave function with the occupations of the electron states,
labelling the delocalized state $e$. The aim is explore the
hybridization of the $X^{2-}$ and $X^{3-}$ excitons, so we
consider 4 electrons and one hole. We take the hole angular
momentum as $+\frac{3}{2}$ (denoted $s_{+\frac{3}{2}}$);
equivalent results are obtained also with $-\frac{3}{2}$. The
electron wave function of the hybridized initial state before
photon emission will have total electron spin $S_e=0$. We
therefore express the ground state as $|i>=A_1|i_1>+A_2|i_2>$,
where:
\begin{eqnarray}
|i_1> & = & \frac{1}{\sqrt{2}}
(|s_\uparrow,s_\downarrow,p_\uparrow,e_\downarrow>-
|s_\uparrow,s_\downarrow,p_\downarrow,e_\uparrow>)|s_{+\frac{3}{2}}>, \nonumber \\
|i_2> & = &
|s_\uparrow,s_\downarrow,p_\uparrow,p_\downarrow>|s_{+\frac{3}{2}}>.\nonumber
\end{eqnarray}
The states $|i_1>$ and $|i_2>$ correspond to the excitons $X^{2-}$
and $X^{3-}$, respectively (fig.~2), and have energies $E_{i_1}$
and $E_{i_2}$. In the Kondo function $|i_1>$, the delocalized
electron ``screens" the net spin in the QD. By diagonalising the
Hamiltonian, we find that the initial ground state has energy
$E_{i}=\frac{1}{2}(E_{i_1}+E_{i_2}-\sqrt{(E_{i_1}-E_{i_2})^2+8V^2})$
and that $A_1=-a/(1+a^2)^{1/2}$, $A_2=-A_1/a$, where
$a=(E_{i_2}-E_{i})/\sqrt{2}V$.

As the energy $E_F$ increases, the ground state evolves from the
$X^{2-}$ to the $X^{3-}$ exciton. The transition occurs when
$E_{i_1}\approx E_{i_2}$ where $E_F\approx E_{1\rightarrow2}=
E^{\rm intra}_2-E^{\rm intra}_1$ where $E^{\rm intra}_1$ and
$E^{\rm intra}_2$ are the intra-dot energies. The hybridization
between $|i_1>$ and $|i_2>$ corresponds to the so-called
mixed-valence regime \cite{book}. To calculate $E_{i}$
numerically, we represent a QD as an anisotropic harmonic
oscillator taking the electron (hole) oscillator frequencies as 25
and 20 (12.5 and 10) meV. Fig.~2a shows $E_{i}$ as a function of
$E_F$.

After photon emission from state $|i>$, the final state $|f>$ has
$S_e=\frac{1}{2}$. The intra-dot configuration in the $X^{2-}$
final state is either a {\it singlet} ($S_{\rm dot}=0$), with
configuration
$\frac{1}{\sqrt{2}}(|s_\uparrow,p_\downarrow>-|s_\downarrow,p_\uparrow>)$,
or a {\em triplet} ($S_{\rm dot}=1$), with configurations
$|s_\uparrow,p_\uparrow>$ and
$\frac{1}{\sqrt{2}}(|s_\uparrow,p_\downarrow>+|s_\downarrow,p_\uparrow>)$
\cite{Nature} (Fig.\ 3a). The PL related to these configurations
are well separated, typically by $\sim 5$ meV
\cite{Nature,Zrenner}, due to the exchange interaction between the
$s$ and $p$ states.
Non-zero optical matrix elements exist for the three final states:
\begin{eqnarray}
|f_1> & = & \frac{1}{\sqrt{6}}(
|s_\uparrow,p_\downarrow,e_\uparrow>+|s_\downarrow,p_\uparrow,e_\uparrow>
-2|s_\uparrow,p_\uparrow,e_\downarrow>), \nonumber
\\ \nonumber |f_2> & = &
\frac{1}{\sqrt{2}}(|s_\uparrow,p_\downarrow,e_\uparrow>-
|s_\downarrow,p_\uparrow,e_\uparrow>) , \nonumber \\
|f_3> & = & |s_\uparrow,p_\uparrow,p_\downarrow>.\nonumber
\end{eqnarray}
In the state $|f_1>$, the intra-dot triplet state with $S_{dot}=1$
is ``screened" by the delocalized electron. $|f_2>$ has the
singlet intra-dot state, and the function $|f_3>$ plays the main
role in the emission of the $X^{3-}$ exciton.

\begin{figure}
\includegraphics[angle=0,width=0.7\textwidth]{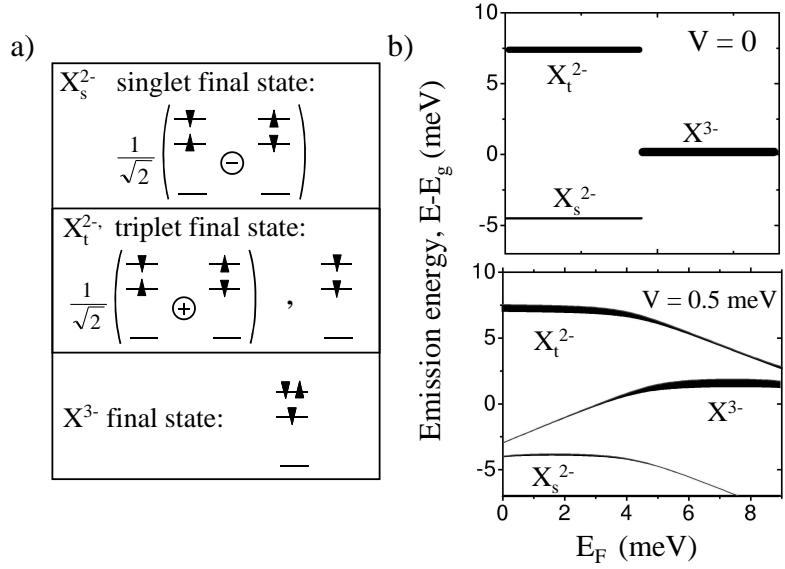}
\caption{\label{fig3} (a) Final intra-dot states related to the
singlet and triplet states of the $X^{2-}$ exciton, and to the
$X^{3-}$ exciton. (b) Calculated energies of the optical transitions
as a function of the delocalized state energy without ($V=0$) and with ($V=0.5$ meV) the
hybridization. The thickness of the line represents the intensity of the emission.}
\end{figure}

The calculated PL spectrum contains three lines (Fig.\ 3), labelled as
$X^{2-}_{t}$, $X^{2-}_{s}$ and $X^{3-}$, with relative intensities
$\frac{3}{4}A^2_1:\frac{1}{4}A^2_1:A^2_2$.
Without the tunnel interaction, there is an abrupt jump in the PL from $X^{2-}$ to $X^{3-}$,
but with the tunnel interaction, the PL shows very clearly hybridization effects.
In particular, in the mixed-valence regime
$E_F\approx E_{1\rightarrow2}$, the intensity of the
$X^{2-}$ lines rapidly decrease and the $X^{3-}$ line appears but the energies of all three
lines depend on the delocalized state energy
$E_F$. Hence, a clear prediction is that in the hybridization regime, the PL
depends on the Fermi energy and therefore also on the gate voltage.

We focus now on the PL line shapes in the Kondo regime
$E_F<E_{1\rightarrow2}$.
While the zero bandwidth model is adequate for the general picture of
the PL, we have to employ the {\em finite} bandwidth model in order to calculate the PL line shapes.
The initial Kondo state is the exciton $X^{2-}$ coupled with the Fermi sea. A trial
function for the $S_e=0$ ground state is \cite{book,trial}:
\begin{eqnarray}
\label{WF1} |i> & = & [A_0|\phi_0>+
\sum_{k>k_F}A_k|\phi_k>]*|s_{+\frac{3}{2}}>; \\
|\phi_0> & = & |s_\uparrow,s_\downarrow,p_\uparrow,p_\downarrow;\Omega>, \nonumber \\
|\phi_k> & = & \frac{1}{\sqrt{2}}\left(
\hat{c}^+_{k,\downarrow}|s_\uparrow,s_\downarrow,p_\uparrow;\Omega>
-\hat{c}^+_{k,\uparrow}|s_\uparrow,s_\downarrow,p_\downarrow;\Omega>\right),
\nonumber
\end{eqnarray}
as represented diagrammatically in Fig.\ 4a.
The configurations are described with the localized states and with the symbol $\Omega$
which denotes all states of the Fermi sea with
$k<k_F$, where $k_F$ is the Fermi momentum.
From the Anderson Hamiltonian, we find that \cite{book}:
\begin{displaymath}
\label{WF2} A_k=-\frac{\sqrt{2} V A_0}{E_1^{intra}+E_k-E_i}, \;
A_0=\frac{1}{(1+2\Delta/\pi\delta)^{1/2}},
\end{displaymath}
where $\Delta=\pi V^2\rho$ ($\rho$ is the 2D DOS). We write the
ground state energy as $E_i= E_1^{\rm intra}+E_F-\delta$, where
$\delta$ is the lowering of energy due to the Kondo effect. If
$\delta<E_{1\rightarrow2}-E_F$, we obtain
\begin{equation}
\delta=(D-E_F)\exp \left(-\frac{\pi(E_2^{\rm intra}-E_1^{\rm intra}-E_F)}{2\Delta}\right)=k_B T_{K}.
\end{equation}
Here, the energy $\delta$ plays the role of the Kondo temperature
in the initial state, $T_{K}$. The temperature $T_{K}$ can be as
high as $\sim 7-14$ K for realistic parameters $E_F=2-3.6$ meV,
$\Delta=1$ meV, and $D\sim30$ meV.

\begin{figure}
\includegraphics[angle=0,width=0.7\textwidth]{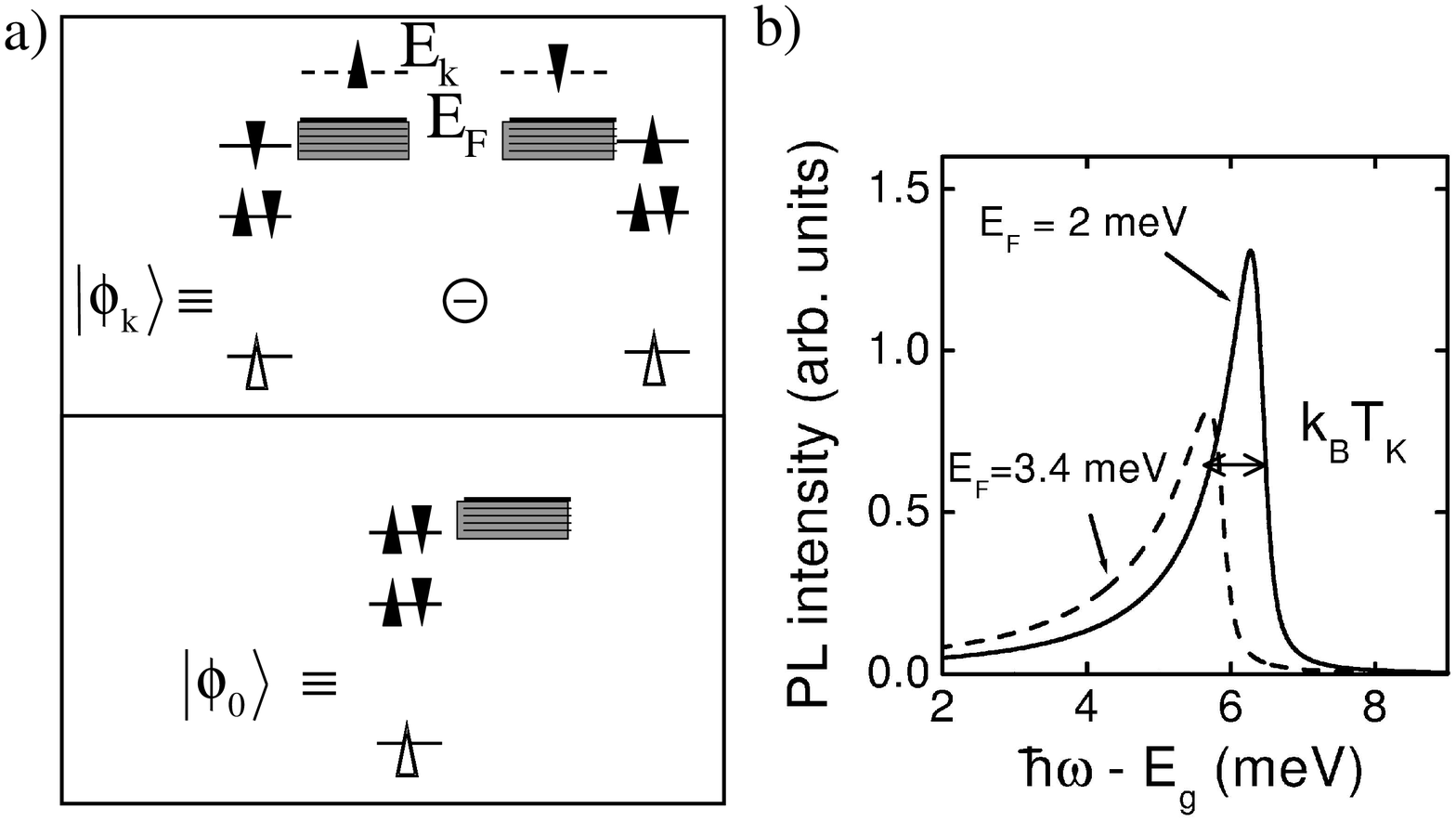}
\caption{\label{fig4} (a) Contributions to the initial state
(\ref{WF1}); (b) Calculated emission spectrum of the $X^{2-}$
exciton with the triplet final state configuration;
$\Delta=1~meV$.  The Fermi energies of $2$ and $3.6~meV$
correspond to $KT_{K}=7$ and $14~K$, respectively.}
\end{figure}

In the regime $E_F<E_{1\rightarrow2}$, the final {\em intra-dot}
configuration is either a singlet or a triplet state (fig.~3a), as
in the case of the zero bandwidth calculation. The triplet state
couples with the Fermi gas, leading to a state with total electron
spin $S_e=\frac{1}{2}$. However, the Kondo temperature for this
final state turns out to be much less than $T_{K}$ allowing us to
use the single-particle final states to calculate the PL spectrum.
The singlet final state, and also the $X^{3-}$ final state, have
higher energy and are broadened by energy $\Delta$ through
tunnelling into empty delocalized states. From state $|i>$ in eq.\
\ref{WF1}, we write the spectral function $I(\omega)$ in the form:
\begin{eqnarray}
\nonumber
I(\omega) & = & -Re[i\sum_{\beta,\beta'}A_\beta
A_{\beta'}F_{\beta,\beta'}],
\end{eqnarray}
where $\beta$ can be either $0$ or $k$, $F_{\beta\beta'} =
<\phi_{\beta}|\hat{V}_{o}^+\hat{R}\hat{V}_{o}|\phi_{\beta'}>$,
and $\hat{R}=1/(\hat{H}+\hbar\omega-E_{i}-i0)$. We find that the
$X^{2-}_{t}$ PL line has an asymmetric shape (Fig.\ 4):
\begin{eqnarray}
X^{2-}_{t}(\omega) & = & V_{opt}^2 \frac{3\Delta A_0^2}{2\pi}
\int_0^{D-E_F}d\epsilon\frac{1}{(\epsilon+k_B
T_{K})^2}Re\frac{-i}{\hbar\omega-E'(X^{2-}_{t})+\epsilon-i\gamma},
\end{eqnarray}
where $E'(X^{2-}_{t})=E(X^{2-}_{t})-k_B T_{K}$ is the renormalized
emission energy and $\gamma$ describes the broadening of the final
state. We can expect $\gamma$ to be small since the relaxation of
the final state requires a spin-flip \cite{Nature}, and in this
case, $k_B T_{K}\gg\gamma_t$, PL line width is equal to $KT_{K}$.
In other words, the PL reflects the spectral DOS near the Fermi
level in the initial Kondo state.

The important result is that in the Kondo regime, both the PL peak
position and the PL line shape depend on the Fermi energy and,
hence, on the gate voltage. Furthermore, as the temperature
increases, the peak in the spectral DOS diminishes rapidly
\cite{book} and therefore the line width of the Kondo-exciton
$X^{2-}_t$ should also be strongly temperature dependent.

In the general case of arbitrary $E_F$, the finite bandwidth model
leads to the PL spectrum similar to that in fig.~3b
\cite{govorov}. The $X^{2-}_s$ and $X^{3-}$ PL lines are close to
Lorentzians, with line widths of $\Delta$ due to the finite
lifetime of the final states. It is important to emphasize that
the $X^{2-}_t$ final state lies at lower energy and does not
suffer from tunnel broadening such that the line shape and line
width are determined by many-body effects.

In summary, we have described novel excitons which can exist
in  self-assembled QDs when there is an interaction between a
charged exciton localized on the dot and delocalized electrons in the wetting layer.
The charged exciton with nonzero
spin is surrounded by a ``cloud" (of radius $\propto (k_B T_{K})^{-1}$) of Fermi electrons.
We predict several striking
manifestations of these novel states in the emission spectra: at low temperature,
the optical lines are strongly dependent on vertical bias and their widths are determined
by the Kondo temperature.

We gratefully acknowledge financial support from the CMSS program at
Ohio University and from the Volkswagen-Stiftung.
One of us (K.K.) gratefully aknowledge J.\ von Delft and A.\ Rosch for
enlightening discussions on Kondo physics.


\begin{thebibliography}{10}
\bibitem{D}
The matrix element $V_k$ decreases with kinetic energy
$E_k$ since the wave functions of delocalized electrons
oscillate. The cut-off parameter $D$ can be estimated as
$\hbar^2/m^*L^2$, where $L$ and $m^*$ are the QD lateral size and
electron mass, respectively.
\bibitem{govorov} More details related to the finite bandwidth model will
be published elsewhere.
\end{thebibliography}

\end{document}